\documentstyle[aps]{revtex}

\begin{document}
\title{
       ROC Analysis and a Realistic Model of
       Heart Rate Variability
}

\author{
    Stefan Thurner,$^1$
  \footnote{Now at: Institut f\"ur Kernphysik, TU Vienna, Austria, 
            e-mail: thurner@kph.tuwien.ac.at}
    Markus C. Feurstein,$^1$
  \footnote{Now at: Institut f\"ur Informationsverarbeitung und 
             Informationswirtschaft, WU Vienna, Austria, \\
           e-mail: Markus.Feurstein@wu-wien.ac.at} 
                     and Malvin C.~Teich$^{1,2}$ 
  \footnote{e-mail:teich@bu.edu} \\
    $^1$ {\it Department of Electrical and Computer Engineering,} 
    {\it Boston University, } {\it  Boston, Massachusetts 02215, USA} \\
    $^2$ {\it Departments of Biomedical Engineering and Physics,}
    {\it Boston University, } {\it  Boston, Massachusetts 02215, USA} \\
}

\maketitle

\begin{abstract}


\noindent
We have carried out a pilot study on a standard collection of
electrocardiograms from
patients who suffer from congestive heart failure,
and subjects without cardiac pathology, using
receiver-operating-characteristic (ROC) analysis.
The scale-dependent wavelet-coefficient standard deviation
$\sigma_{{\rm wav}}(m)$, a
multiresolution-based analysis measure, is found to be superior to
two commonly used measures of cardiac dysfunction when the
two classes of patients cannot be completely separated.
A jittered integrate-and-fire model
with a fractal Gaussian-noise kernel provides a realistic
simulation of heartbeat sequences for both heart-failure patients
and normal subjects.
\\
PACS number(s): 87.10.+e, 87.80.+s, 87.90.+y
\end{abstract}

\vspace{0.5cm}

{\it Introduction.}--
The interbeat-interval (R-R) time series
of the human heart exhibits scaling behavior,
as evidenced by the power-law form of its power spectrum which
decreases as
$f^{-{\delta}}$ for sufficiently low frequencies $f$
\cite{BASS94,KOBA82}.
However, other features
associated with physiological
markers \cite{GORD81} are also present in the power spectrum
at particular frequencies.
Moreover, it is well known that the heartbeat
time series is nonstationary, reflecting biological adaptability.

Multiresolution wavelet
analysis \cite{DOBE92,MALL89,MEYE86,ALDR96,METN97} provides
an ideal means of decomposing a signal into its components at different
scales, and at the same time has the salutary effect of
eliminating nonstationarities
\cite{ARNE95,ABRY96,TEIC96}.
We previously carried out a study \cite{THUR97B}
in which wavelets were used to analyze
the sequence of interbeat intervals from 
a standard electrocardiogram (ECG) database \cite{data}.
Using the wavelet-coefficient standard deviation
$\sigma_{{\rm wav}}(m)$, where $m$ is the scale, we
discovered a critical scale window
over which it was possible to perfectly discriminate
heart-failure patients from normal subjects.
The presence of this scale window has
been confirmed in a recent Israeli-Danish
study of diabetic patients who had not yet developed clinical signs of
cardiovascular disease \cite{ASHK98}. These two studies, in conjunction
with earlier investigations involving the {\it counting} statistics of the
heartbeat (as opposed to the time-{\it interval} statistics considered here)
\cite{TURC93,TURC96,TEIC96B}, have led us to the conclusion
that scale-dependent measures (such as the wavelet-coefficient standard
deviation)
outperform scale-independent ones (such as the scaling exponent $\delta$)
in discriminating patients with cardiac dysfunction from
normal subjects.

The perfect separation achieved in our initial study
endorsed the choice of
$\sigma_{{\rm wav}}(m)$ as a measure of importance,
at least for $m = 5$ (corresponding to $2^5$ = $32$ heartbeat
intervals). The results of
most studies are
seldom so clear-cut, however.
In circumstances where there is incomplete separation between
two classes of subjects, as observed for other measures using
these identical data sets \cite{PENG93,PENG95}, or in applying
our measure to large collections of out-of-sample
data sets, the relative abilities of different measures
in determining the presence of disease is best established by the use of
receiver-operating-characteristic (ROC)
analysis \cite{SWET88}. ECG recordings of reduced duration
also give rise to incomplete separation using our wavelet measure.

In this Letter we use ROC analysis to
quantitatively compare the tradeoff between data length and
discriminability provided by $\sigma_{{\rm wav}}(m)$, and by two
other widely used heart rate variability measures of cardiac dysfunction.
We then 
develop a mathematical model for heartbeat time-series generation for both
heart-failure patients and normal subjects.

{\it Using ROC analysis to identify cardiac dysfunction.}--
We wish to establish the tradeoff
between reduced data length on one hand, and misidentifications
(misses and false positives) on the other. The ROC curve is a
plot of sensitivity {\it vs} specificity as the threshold
parameter is swept; its use requires no assumptions about the
statistical nature of the data. The area under the ROC curve serves
as a well-established index of diagnostic
accuracy \cite{SWET88}; the minimum value (0.5) arises from assignment
by pure chance whereas the maximum value (1.0) corresponds to perfect
assignment. Carrying out the ROC calculations permits us to quantitatively
compare the abilities of different measures in diagnosing disease.

In Fig. 1 we present ROC areas, as a function of
data length, using the three different measures. The solid
curve in Fig. 1(a) shows ROC area when discriminability is
determined by the
wavelet-coefficient standard deviation $\sigma_{\rm wav}(m=5)$,
using Daubechies 10-tap wavelets \cite{DOBE92}.
Figure 1(b) represents the area when
the interbeat-interval standard deviation
$\sigma_{\rm int}$ is used instead. The importance of
this measure has long been known
\cite{WOLF78} and it is now commonly used in heart rate
variability analysis \cite{MALI96}.
Figure 1(c) provides the ROC area for yet another well-known measure,
the spectral scaling exponent $\delta$
estimated at ultra-low ($< 0.003$ Hz) and at very-low ($< 0.04$ Hz)
frequencies \cite{KOBA82,MALI96}.

It is clear from Fig. 1 that $\sigma_{\rm wav}(m=5)$ is
the only measure of the three that, for the ECG recordings in our pilot
study, ever achieves an ROC area of unity, thereby indicating perfect ability
to separate the heart-failure patients from the normal subjects, and it
does so
with as few as $20\,000$ heartbeats (corresponding to 4 or 5 hours of data).
It is equally evident from Fig. 1 that the measure of choice for
ECG recordings with fewer than $3\,500$ heartbeats (corresponding to
about 45 minutes
of data) is not $\sigma_{\rm wav}(m=5)$ but rather
$\sigma_{\rm int}$.
This transition occurs because
$\sigma_{\rm int}$ depends only on the short-term behavior of the R-R sequence
\cite{THUR97B,TURC96} whereas the wavelet measure depends on both the short-
and long-term behavior. It is also apparent from Fig. 1 that
$\sigma_{{\rm wav}}(m=5)$ and $\sigma_{\rm int}$
{\it always} outperform the scaling exponent $\delta$, whatever the data
length. Moreover, because $\delta$ reflects
the long-duration properties of the interbeat-interval sequence
\cite{THUR97B,TURC96}, its error brackets
become unacceptably large as data length decreases (see Fig. 1(c)) so
that it can only be reliably calculated for long data sets.
Since we have previously shown \cite{TURC96} that the spectrum
of the R-R sequence, the spectrum of the generalized rate, and the
Allan factor all exhibit scaling exponents that are similar in value
(denoted $\delta$, $\beta$, and $\gamma$,
respectively, in \cite{TURC96}), the use of
scale-dependent measures such
as $\sigma_{{\rm wav}}(m=5)$ and $\sigma_{\rm int}$ is likely to prove
superior
to the use of scale-independent
measures such as $\delta$, $\beta$, or $\gamma$.

{\it Generating a realistic heartbeat sequence.}--
The generation of a mathematical point process that faithfully emulates the
human heartbeat could be of importance in a number of venues, including
application to pacemaker excitation.
Integrate-and-fire (IF) models, which are physiologically plausible, have been
developed for use in cardiology \cite{HYND75,BERG86}. In the paper
by Berger et al. \cite{BERG86},
for example, an integrate-and-fire model was constructed by
integrating an underlying rate
function $R(t)$ until it reached a fixed threshold $\theta$,
whereupon a point event was triggered and the integrator
reset. The occurrence time for the $(k+1)$st beat is then
implicitly given by
$\theta=\int_{t_k}^{t_{k+1}} \, R(\tau) \, d\tau$.
Modeling the stochastic component of the rate
function as band-limited
fractal Gaussian noise (FGN), which
introduces scaling behavior into the
heart rate, and setting $\theta=1$, results in
improved agreement with experiment \cite{TURC96}.
This fractal-Gaussian-noise
integrate-and-fire (FGNIF) process
requires four parameters: the scaling exponent,
the relative strength of the FGN spectrum, and
lower and upper limits for the noise band.
The FGNIF has been quite successful in fitting a whole host of
interval- and count-based measures of the heartbeat sequence for
both heart-failure patients and normal subjects \cite{TURC96}. However, it
is not able to accommodate the differences observed in the behavior
of $\sigma_{\rm wav}(m)$ for the two classes of data.

To remedy this defect, we have constructed
a jittered version of this model which we dub
the FGNJIF.
The process is generated as follows. Preliminary event
occurrence times $t_i^{\rm pri}$ are generated by the FGNIF;
a Gaussian jitter distribution of standard deviation $J$ is then convolved
with each of the $t_i^{\rm pri}$ to determine the
times of the final points $t_i$ \cite{THUR97A}.
Increasing the jitter parameter imparts additional randomness
to the R-R time series at small scales,
thereby increasing  $\sigma_{\rm wav}$ at small values of $m$ and,
concomitantly, the power spectral density at large values of the
frequency $f$.

In Fig. 2, we present simulations for the wavelet-coefficient standard
deviation
$\sigma_{{\rm wav}}^{\rm sim}$ versus scale $m$
using the FGNJIF model. For $J$ = 0 (circles), the results
reduce to those for the FGNIF model, and the shape of the simulated
curves is in reasonably good accord with
experimentally observed curves for normal subjects (Fig. 3(b) left).
As the jitter standard deviation $J$ increases above 0, the curves
in Fig. 2 bend upward
at small values of the scale $m$, and the curves begin to match those
for heart-failure patients (Fig. 3(b) right). The
increased jitter also gives rise to a whitening of the spectrum
at high frequencies, as expected,
so that the distinctions in the spectra between
heart-failure patients and normal subjects
are properly mimicked by the FGNJIF model
(compare Fig. 3(d) right and left).
Typical values of $J$ that accommodate the data lie in the range 0.01
to 0.06 for heart-failure patients and in the range
0 to 0.02 for normal subjects.

The input parameters
used to generate some of the simulated curves displayed in Fig. 2, as
well as estimates of the quantity $\alpha$ obtained from $\sigma_{{\rm
wav}}^{\rm sim}$
over different ranges of $m$ \cite{slope}, are presented in
the lower portion of Table 1. Care should be exercised in referring to
$\alpha$, however, since it is sometimes estimated over a very narrow
region of $m$ and therefore cannot be considered as a slope or
scaling exponent. Examination of Figs. 2 and
3(b) (right) reveals that the heart-failure 
wavelet-coefficient standard deviation curves
gently bend upward at small scales. It is clear from the lower portion of
Table 1 that
increasing the jitter at the input to the simulation ($J_{{\rm inp}}$)
results in a
reduction in $\alpha(1 \leq m \leq 3)$,
mimicking the observations for heart-failure patients as
shown in the upper portion of Table 1.
A related distinction for the two classes of patients was observed by
Peng {\it et al.} \cite{PENG95}
using detrended fluctuation analysis but, in contrast to us, those
authors literally speak of two scaling regions.
The effective scaling exponents $\alpha(3 \leq m \leq 10)$
in the large-scale (low-frequency) regime,
slowly decrease with increasing $J_{{\rm inp}}$, as expected, since the curves
bend up at low scales.

In Fig. 3 we demonstrate that the FGNJIF simulation does a
rather remarkable job of reproducing the actual data for a number of
key measures used in heart rate variability analysis. The results
of the model calculations
are compared with ECG data for a single normal subject
(left column) and for a single heart-failure patient
(right column).
The input
parameters (mean interbeat interval $\langle \tau_i \rangle_{\rm inp}$
and wavelet scaling exponent $\alpha_{\rm inp}$)
used in the model were obtained from the data sets.
The jitter standard deviation used for this particular normal
simulation was zero whereas it was $J=0.023$ for the heart-failure
simulation.

Finally, it is of interest to examine the global performance
of the fractal-Gaussian-noise
jittered integrate-and-fire (FGNJIF) model for the
entire collection of data sets in
our pilot study. To this end we construct a simulated ROC curve using
the measure $\sigma_{{\rm wav}}^{\rm sim}(m=5)$. The results for
the underlying area are
presented as the dashed curve in Fig. 1(a). It
is derived from 27 simulations \cite{THUR97A}, each
with $70\,000$ events, to match the pilot-study data. Model
parameters were determined from the actual ECGs and
the scaling exponent was estimated from the slope of the
wavelet-coefficient standard deviation over the range
$(3 \leq m \leq 10)$ \cite{slope}. The jitter standard deviation value $J$
was established by finding the best fit to $\sigma_{{\rm wav}}(m)$.
The corner frequency
was taken to be 0.0005 times the mean rate of the process (simply
0.0005 for the simulation) as suggested by the results of Ref.
\cite{THUR97A}.

The global simulation in Fig. 1(a) (dashed curve) follows the trend
of the data (solid curve) quite nicely, but there is room for
improvement since it falls short of the agreement of the
individual simulations illustrated in Fig. 3.
It will be of interest to consider modifications of the
model, including the possibility of
nonlinear dynamical behavior \cite{POON97}, that might bring the
simulated ROC curves
into better accord with the data-based curves.


\begin{table}[h]
\begin{center}
\begin{tabular}{|l|c|c|c|c|c|c|c|}
\multicolumn{8}{|c|} {\bf Interbeat Intervals  (Pilot Study) } \\
\hline
\multicolumn{1}{|l}{Class} &
\multicolumn{3}{c||}{} &
\multicolumn{4}{c|}{Experimental Results} \\
\hline
\multicolumn{1}{|c}{} &
\multicolumn{1}{c}{} &
\multicolumn{1}{c}{} &
\multicolumn{1}{c||}{} &
\multicolumn{1}{c|}{$\langle \tau_i \rangle$} &
\multicolumn{1}{c|}{$\alpha  (1 \leq m \leq 10)$} &
\multicolumn{1}{c|}{$\alpha  (1 \leq m \leq 3)$ } &
\multicolumn{1}{c|}{$\alpha  (3 \leq m \leq 10)$ }\\
\hline
\multicolumn{1}{|l}{Normal Subjects (12)} &
\multicolumn{1}{c}{} &
\multicolumn{1}{c}{} &
\multicolumn{1}{c||}{} &
\multicolumn{1}{c|}{ $0.79 \pm 0.08$} & 
\multicolumn{1}{c|}{ $1.23 \pm 0.13$} & 
\multicolumn{1}{c|}{ $1.40 \pm 0.37$} &
\multicolumn{1}{c|}{ $1.22 \pm 0.11$} \\
\multicolumn{1}{|l}{H-F Patients (15)} &
\multicolumn{1}{c}{} &
\multicolumn{1}{c}{} &
\multicolumn{1}{c||}{} &
\multicolumn{1}{c|}{$0.67 \pm 0.13$} & 
\multicolumn{1}{c|}{$1.35 \pm 0.22$} & 
\multicolumn{1}{c|}{$0.26 \pm 0.60$} &
\multicolumn{1}{c|}{$1.57 \pm 0.17$} \\
\hline
\hline
\multicolumn{8}{|c|} {\bf Interbeat Intervals (Simulations) } \\
\hline
\multicolumn{1}{|l}{Class} &
\multicolumn{3}{c||}{Input Parameters} &
\multicolumn{4}{c|}{Simulation Results} \\
\hline
\multicolumn{1}{|c}{} &
\multicolumn{1}{c|}{$\langle \tau_i \rangle_{\rm inp}$} &
\multicolumn{1}{c|}{$\alpha_{\rm inp}$} &
\multicolumn{1}{c||}{$J_{\rm inp}$} &
\multicolumn{1}{c|}{$\langle \tau_i \rangle$} &
\multicolumn{1}{c|}{$\alpha (1\leq m\leq 10)$} &
\multicolumn{1}{c|}{$\alpha (1\leq m\leq 3)$ } &
\multicolumn{1}{c|}{$\alpha (3\leq m\leq 10)$ }\\
\hline
\multicolumn{1}{|l}{Normal}  & 
\multicolumn{1}{c|}{0.8}   &  
\multicolumn{1}{c|}{1.1}   & 
\multicolumn{1}{c||}{0.00} & 
\multicolumn{1}{c|}{0.78}  &  
\multicolumn{1}{c|}{1.40}  & 
\multicolumn{1}{c|}{2.03}  & 
\multicolumn{1}{c|}{1.31}  \\
\multicolumn{1}{|l}{Heart-Failure} & 
\multicolumn{1}{c|}{0.7} &  
\multicolumn{1}{c|}{1.4} & 
\multicolumn{1}{c||}{0.00}& 
\multicolumn{1}{c|}{0.70} &  
\multicolumn{1}{c|}{1.67} & 
\multicolumn{1}{c|}{2.07} & 
\multicolumn{1}{c|}{1.60} \\
\multicolumn{1}{|c}{} &
\multicolumn{1}{c}{}  &
\multicolumn{1}{c}{}  &
\multicolumn{1}{c||}{0.01} & 
\multicolumn{1}{c|}{0.70} &  
\multicolumn{1}{c|}{1.54} & 
\multicolumn{1}{c|}{1.28} & 
\multicolumn{1}{c|}{1.58}  \\
\multicolumn{1}{|c}{} &
\multicolumn{1}{c}{}  &
\multicolumn{1}{c}{}  &
\multicolumn{1}{c||}{0.05} & 
\multicolumn{1}{c|}{0.70} &  
\multicolumn{1}{c|}{1.10} & 
\multicolumn{1}{c|}{0.16} & 
\multicolumn{1}{c|}{1.35} \\
\multicolumn{1}{|c}{} &
\multicolumn{1}{c}{}  &
\multicolumn{1}{c}{}  &
\multicolumn{1}{c||}{0.10} & 
\multicolumn{1}{c|}{0.70} &  
\multicolumn{1}{c|}{0.84} & 
\multicolumn{1}{c|}{0.04} & 
\multicolumn{1}{c|}{1.10}  \\
\end{tabular}
\end{center}
\caption{ The upper portion of the table presents mean interbeat intervals
and quantities $\alpha$ (with their standard deviations) 
estimated from the wavelet-coefficient standard deviation for the 12 
normal subjects and 15 heart-failure patients comprising our pilot study,
collected in two separate groups. Three heart failure patients
who also suffer from atrial fibrillation are included
among the heart-failure patients.
Estimates of the effective values of $\alpha$
are determined over the entire range of $m$, and also over two
smaller subregions: $1\leq m\leq 3$
(the values of $\alpha$ in this region are not scaling
exponents -- see text) and $3\leq m\leq 10$ 
[26].
There is a significant difference in the behavior of $\alpha$ for
the two classes of patients.
The lower portion of the table provides estimates obtained from
the FGNJIF model. It is clear that the jitter standard deviation
$J$ plays a crucial role in accommodating the heart-failure
observations.  }
\label{tab1}
\end{table}


\newpage
\begin{figure}
\caption{ Diagnostic accuracy (area under ROC curve) vs data length (number
of heartbeats).
A maximum area of unity corresponds to the correct assignment of each
patient to the
appropriate class. The solid curve in the
upper panel (a) is obtained by using the wavelet coefficient
standard deviation at scale 5 (similar results are obtained at scale 4);
the middle panel (b) arises from using the interbeat-interval standard
deviation; and the lower panel (c) emerges when using
the spectral scaling exponent. The areas
are based on averages of the first 10 data segments for 64, 128, 512, 1024,
3500, and 7000 events (the leftmost six data points in (a) and (b)),
and on 5, 3, 2, and 1 segments of $14\,000$, $20\,000$, $35\,000$, and
$70\,000$ events, respectively (the rightmost 4 data points).
$\sigma_{\rm wav}$
is the only measure of the three that achieves 100\% sensitivity at
100\% specificity for the data in our pilot study, and it does so with as
few as
$20\,000$ heartbeats (corresponding to 4 or 5 hours of data). For data lengths
less than $3\,500$ events (corresponding to about 45 minutes of data),
the best performance is provided by
 $\sigma_{\rm int}$ rather than  $\sigma_{\rm wav}$.
The dashed curve in (a) is derived from 27 simulations of the
fractal-Gaussian-noise jittered integrate-and-fire (FGNJIF) model (see
text).
 }
\end{figure}

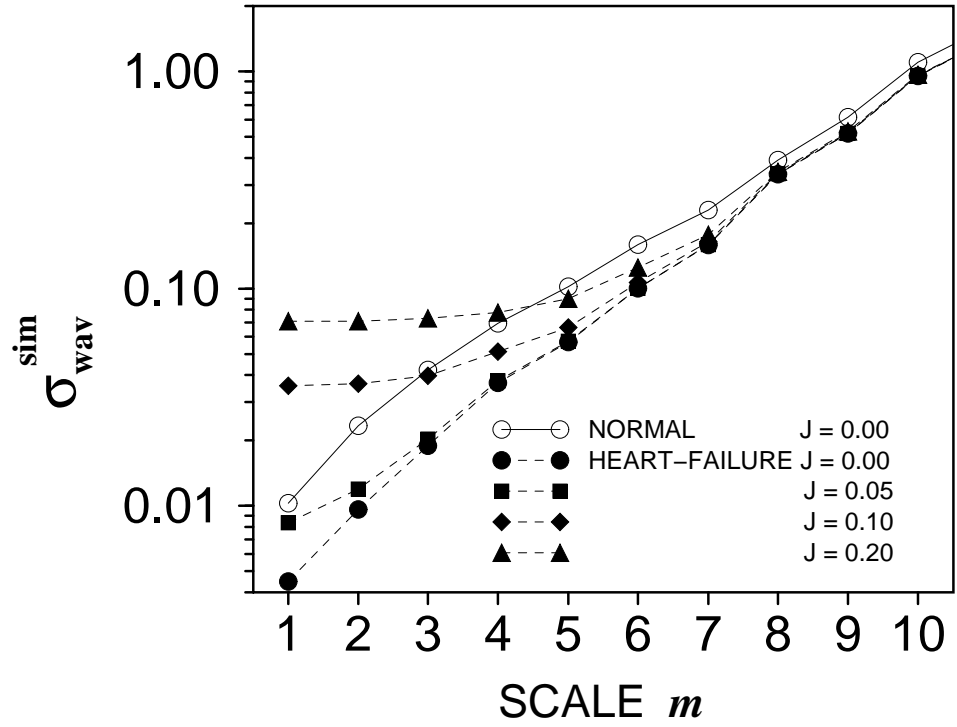
\begin{figure}
\caption{ Simulated wavelet-coefficient  standard deviation
          $\sigma_{{\rm wav}}^{\rm sim}$ curves versus scale $m$
          using the FGNJIF model. For $J$ = 0,
          the results reduce to those for the FGNIF model.
          As the jitter standard deviation $J$ increases from 0,
          the curves bend up at low values of $m$,
          accurately mimicking the results for
          heart-failure patients as is evident from Fig. 3(b)
          right.
}
\end{figure}

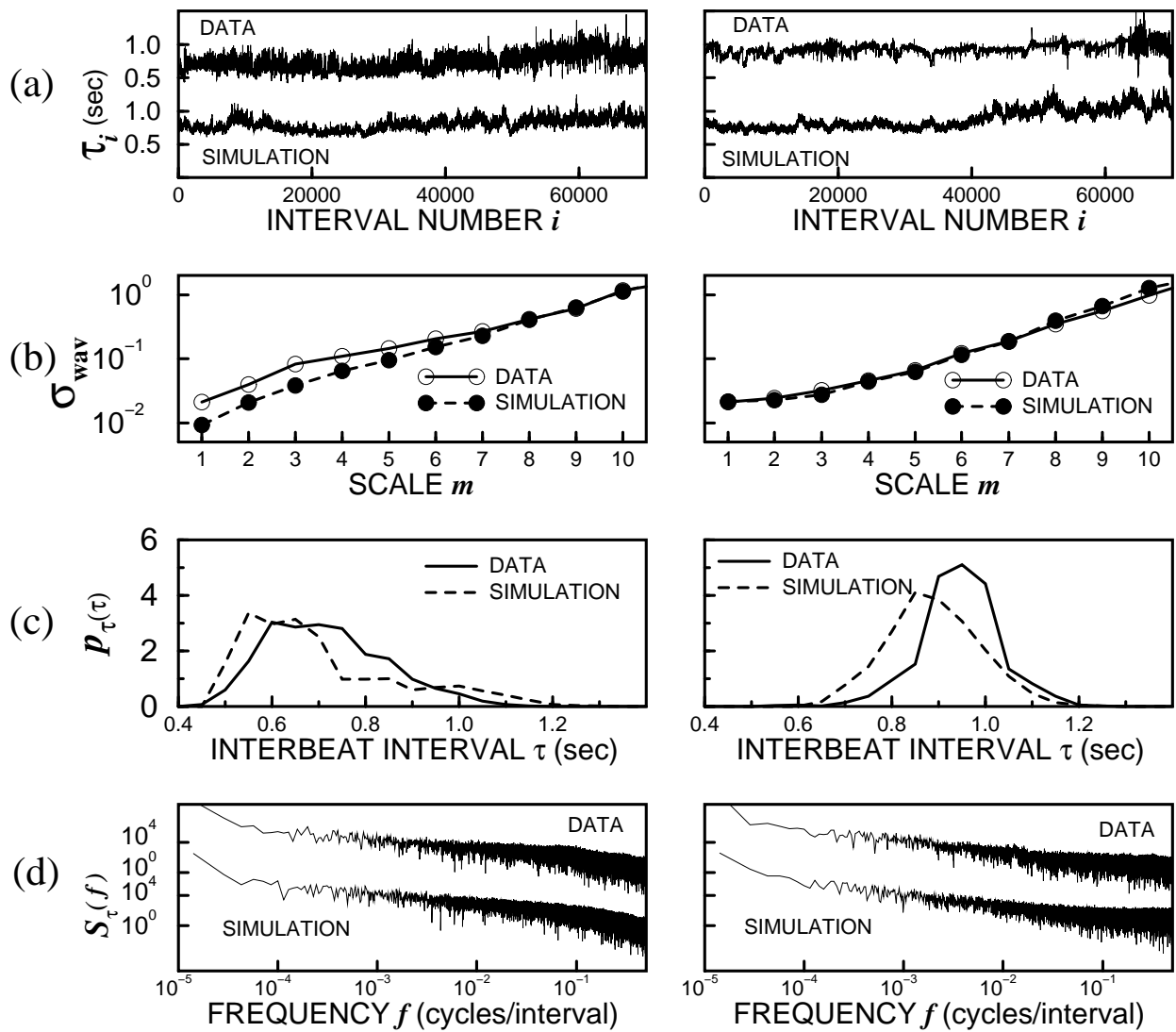
\begin{figure}
\caption{ Comparison of data from a single normal subject
         (data set 16265, left column), and a single heart-failure patient
         (data set 6796, right column), with results obtained using the
         FGNJIF simulation. The
         parameters $\langle \tau_i \rangle $ (representing the mean
         interbeat interval) and $\alpha$ (representing the scaling exponent)
         used in the simulation were obtained from the actual data sets.
         Good fits were obtained by using a jitter
         parameter $J=0$ for the normal subject and
         $J=0.023$ for the heart-failure patient.
         (a) R-R interval sequence over the entire data set.
         Qualitative agreement is apparent in both cases.
         (b) Wavelet-coefficient standard deviation {\it vs} scale.
         The model reproduces the
  scaling properties of the data in both cases, particularly the
         gentle increase of $\sigma_{{\rm wav}}$ at small scale values.
         (c) Interbeat-interval histogram.
         The model captures the narrowing of the histogram
         (reduction of $\sigma_{{\rm int}}$) for heart failure.
         (d) Spectrum of the sequence of R-R intervals. The simulation
         captures the subtle distinctions quite well, including
         the whitening of the heart-failure spectrum at high frequencies.
}
\end{figure}


\begin{thebibliography}{99}
\bibitem{data}
The R-R recordings were drawn from
the Beth-Israel Hospital (Boston, MA) Congestive Heart-Failure
Database comprising 12  records
from normal patients  (age: 29-64 years, mean 44 years)
and 15 records from severe congestive heart-failure patients (age 22-71
years, mean 56 years).
The recordings, which form a standard database for evaluating the merits of
various measures for identifying heart failure,
were made with a Holter monitor,  digitized at a fixed value of 250 Hz.
Three of the 15 heart-failure patients also suffered  from atrial
fibrillation.
Detailed characterization of the data sets is presented in Table 1 of Ref.
\cite{TURC96}.
We used the first $M =  70\,000$ interbeat intervals (total time
duration $T \approx $ 20 h) of each of these records.
\bibitem{slope}
In the region of large $m$,
the scaling exponent $\alpha$ is estimated from an individual
wavelet-coefficient standard-deviation curve (as
shown in Fig. 2 or Fig. 3(b)) by calculating the slope of its square
(rendering it a variance so that it corresponds to other standard
scaling-exponent measures) on a base-10 log-log plot. Thus
$\alpha = [d({\rm log_{10}}\sigma_{{\rm wav}}^{2}(m))]/[d({\rm
log_{10}}2^m)] =
[2/{\rm log_{10}}2][d({\rm log_{10}}\sigma_{{\rm wav}}(m))/dm] =
6.644[d({\rm log_{10}}\sigma_{{\rm wav}}(m))/dm]$.
\end{thebibliography}
\end{document}